\documentclass[aps,prb,twocolumn,superscriptaddress]{revtex4}
\usepackage{times}
\usepackage{amsmath, amssymb}
\usepackage{amsfonts}
\usepackage{graphicx}
\usepackage{bm}
\usepackage{hyperref}
\hypersetup{
        colorlinks=true,
}

\begin{document}

\renewcommand{\ol}[1]{\overline{#1}}
\newcommand{\ket}[1]{|#1\rangle}
\newcommand{\mycomment}[1]{}
\newcommand{\mb}[1]{\mathbf{#1}}
\renewcommand{\ol}[1]{\overline{#1}}
\renewcommand{\vr}{{\bm{r}}}
\newcommand{\vk}{{\bm{k}}}
\newcommand{\dd}{\mathrm{d}}
\newcommand{\ii}{\mathrm{i}}
\newcommand{\Ref}[1]{Ref.\,\onlinecite{#1}}
\renewcommand{\eqref}[1]{Eq.\,(\ref{#1})}
\newcommand{\figref}[1]{Fig.\,\ref{#1}}
\newcommand{\tf}[2]{\theta{{#1} \brack {#2}}}
\newcommand{\pf}{\mathrm{Pf}}

\title{Measuring Modular Matrices by Shearing Lattices}
\author{Yi-Zhuang You}
\affiliation{Department of physics, University of California, Santa Barbara, CA 93106, USA}
\author{Meng Cheng}
\affiliation{Microsoft Research, Station Q, Elings Hall,
University of California, Santa Barbara, California 93106-6105, USA}

\begin{abstract}

A topologically ordered phase on a torus possesses degenerate ground states that transform nontrivially under the modular transformations of the torus, generated by Dehn twists. Representation of modular transformations on the ground states (modular matrices) characterizes the topological order. We show that the modular matrices can be numerically measured as the non-Abelian Berry phase of adiabatic deformations of the lattice model placed on a torus. We apply this method to the example of a gauged $p_x+\ii p_y$ superconductor, and show that the result is consistent with the topological quantum field theory descriptions.

\end{abstract}

\maketitle

When a two-dimensional topologically ordered phase~\cite{Wen90a} inhabits a manifold, it possesses a robust ground state degeneracy~\cite{Wen90b} dependent on the topology of the manifold. The topological ground state degeneracy can not be lifted by any local perturbation up to exponentially small corrections. The topological order (or the long-range entanglement\cite{Chen2010}) in these degenerate ground states can be characterized by the unitary transformations among them induced by the diffeomorphisms of the manifold, which can be described by topological quantum field theories(TQFT) at low energy. Particularly interesting are the ``large'' diffeomorphisms that are not continuously connected to the identity, whose equivalence classes under the identity component form the mapping class group (MCG) of the manifold. It is widely believed that the (projective) representations of the MCG on the topologically-degenerate ground states, defined as the non-Abelian Berry phases of the ground states under adiabatic deformations of the manifold~\cite{Wen1993}, can be used to characterize the topological phase~\cite{Kitaev06a, RSW, Nayak08}.
 
In two dimensions, the MCG of a torus is the modular group $\mathrm{SL}(2, \mathbb{Z})$, generated by the Dehn twist and $\pi/2$ rotation. The matrix representations of the two generators of $\mathrm{SL}(2, \mathbb{Z})$ on the topological ground states, called the modular $S$ and $T$ matrices, are believed to completely characterize the topological phase. For example, one can read off the quantum dimensions of quasiparticles, fusion rules and braiding statistics from the modular matrices. 
 It is therefore an important question to determine the modular matrices from microscopic Hamiltonians or ground state wavefunctions in order to identify the topological order. In continuous space (e.g. quantum Hall systems) Dehn twists can be realized by adiabatically deforming the metric and was exploited to find the Hall viscosity of quantum fluids~\cite{AvronPRL1995, ReadPRB2009,ReadPRB2011}. Recently a number of proposals to extract the modular matrices numerically have appeared~\cite{ZhangPRB2012, TuPRB2012, ZaletelPRL2013, CincioPRL2013, Zaletel_arxiv,  He_arxiv, Moradi_arxiv, ZhangPRB2015}, mostly from entanglement measurement or ground state wavefunction overlap.

\begin{figure}[t!]
	\centering
	\includegraphics[width=0.8\columnwidth]{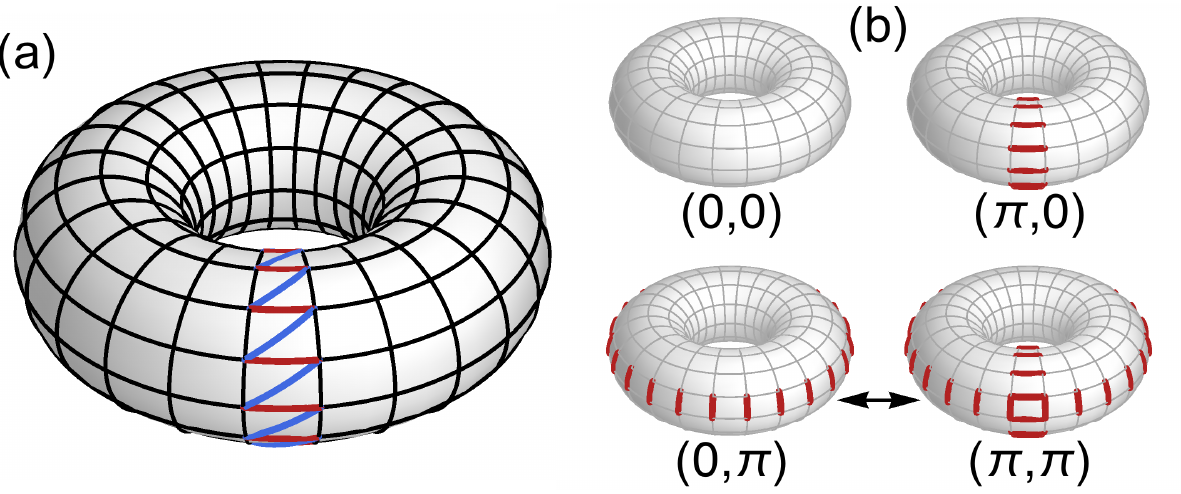}
	\caption{(Colored online.) (a) One-step twist by smoothly weaken the horizontal red bonds and strengthen the diagonal blue bonds. (b) Four flux configurations (boundary conditions) labeled by $(\Phi_x,\Phi_y)$. Red bounds are with minus signs. The configurations $(0,\pi)$ and $(\pi,0)$ are connected by a full Dehn twist along the direction shown in (a).}
	\label{fig:twist}
\end{figure}

 While these proposals have been proven successful in numerical applications~\cite{TuPRB2012, ZhuPRB2013, ZaletelPRL2013, ZhuPRL2014, Mei_arxiv, ZaletelPRB2015}, the entanglement-based methods~\cite{TuPRB2012, ZaletelPRL2013} mostly work in a cylinder geometry, and the methods based on wavefunction overlap~\cite{ZhangPRB2012, CincioPRL2013} do not actually calculate the adiabatic Berry phases. In this work we show that for lattice models on a torus, Dehn twists, or ``shear'' deformation of the system can be directly implemented by adiabatically reconnecting the bonds of the lattice (see \figref{fig:twist}(a) for an illustration of the square lattice).   Generally in a finite-size system (e.g. a $L\times L$ torus ), the resulting adiabatic Berry phase
 $\Theta$ has the following dependence on system size:
\begin{equation}
  \Theta=\alpha+\beta L^2+O(L^{-2}).
\end{equation}
Here $\alpha$ is a universal quantity, determined by the chiral central charge $c_-\,\text{mod }24$ and topological twists of quasiparticles. The Dehn twist $T$ matrix can be obtained by measuring this non-Abelian Berry phase in the ground state manifold.
Once the Dehn twists along both the meridian ($T_y$) and longitudinal ($T_x$) cycles of the torus are found, the $S$ matrix can be derived from the composition of Dehn twists as $S=T_yT_x^{-1}T_y$. We apply this method to the $p_x+\ii p_y$ topological superconductor.

\emph{Dehn twists of $p_x+\ii p_y$ superconductors.}---
We consider a $p_x+\ii p_y$ superconductor with the following Hamiltonian:
\begin{equation}
\begin{split}
  H=&-t\sum_{\langle \vr\vr'\rangle}(c_\vr^\dag c_{\vr'}+\text{h.c.})-\mu\sum_\vr c_\vr^\dag c_\vr \\
  &+\sum_{\langle \vr\vr'\rangle}(\Delta_{\vr,\vr'}c_\vr^\dag c_{\vr'}^\dag +\text{h.c.})
  \end{split}
  \label{eq: H}
\end{equation}
Hereinafter we set the hopping amplitude to $t=1$ as the energy unit, and $\Delta_{\vr,\vr'}=-\Delta_{\vr',\vr}$ is an odd-parity pairing which follows the $p_x+\ii p_y$ pairing symmetry: $\Delta_{\vr,\vr+\mb{e}_x}=\Delta, \Delta_{\vr,\vr+\mb{e}_y}=\ii\Delta$. When $|\mu|<4$, the Hamiltonian describes a topological superconductor with Chern number $\nu=1$, also known~\cite{Read00} as the ``weak pairing'' phase; while in the ``strong pairing'' phase $|\mu|>4$ it describes a trivial superconductor. In the momentum space, the Bogoliubov-de Gennes (BdG)  Hamiltonian reads
\begin{equation}
  \begin{split}
	  H=&\frac{1}{2}\sum_\vk (c_{\vk}^\dag, c_{-\vk}) H_\vk \begin{pmatrix} c_\vk\\ c_{-\vk}^\dag\end{pmatrix},\\
  H_\vk=&[-2(\cos k_x+\cos k_y)-\mu]\sigma_z\\
  &+ \Delta (\sin k_x\sigma_x+\sin k_y \sigma_y).
  \end{split}
  \label{eq: H in k}
\end{equation}
We notice that the $p_x+\ii p_y$ superconductor itself is a short-range entangled state, i.e. the topological entanglement entropy is zero. When put on a torus with periodic boundary conditions, there is a unique ground state and the $S, T$ matrices are both scalars.

However, since this is a fermionic system, one has four choices of boundary conditions for the fermions on a torus: $c_{\vr+L_x}=e^{\ii \Phi_x}c_\vr, c_{\vr+L_y}=e^{\ii \Phi_y}c_\vr$. 
\footnote{Here we notice a potentially confusing convention in defining the boundary conditions of fermions. If we actually consider fermions living on a torus which embeds in $\mathbb{R}^3$, the boundary conditions of $c_\vr$ is shifted relatively to the actual flux by $\pi$, because of the spin connections on the torus.}
Here $\Phi_{x,y}$ is the amount of magnetic flux threading through the holes of the torus and the distinct values are $0$ and $\pi$ mod $2\pi$ due to the flux quantization. We notice that in a topological superfluid with gapless Goldstone modes, the phase $\theta$ of the pairing order parameter(i.e. $\Delta=|\Delta|e^{i\theta}$) has winding numbers $w_x$ and $w_y$ around the $x$ and $y$ directions of the torus due to the magnetic fluxes. Therefore, the ground state energy receives the following contribution from the phase stiffness:
	$\frac{\rho_\text{s}}{2}\int d^2\vr\,(\nabla\theta)^2=2\pi^2\rho_\text{s}(w_x^2+w_y^2)$,
where $\rho_\text{s}$ is the superfluid density per unit area.
As a result, the four sectors corresponding to $(w_x,w_y)=(0,0), (1,0), (0,1), (1,1)$ are not degenerate, and have the energy splitting independent of the system size~\cite{Bonderson12d}. To avoid this issue, one can couple the fermions to a dynamical 2D gauge field, then the Anderson-Higgs mechanism will completely gap out the Goldstone mode. In our calculation, the order parameter is treated as a static extrinsic parameter, so the phase fluctuations do not enter.

For a $p_x+\ii p_y$ superconductor on an $L_x\times L_y$ lattice, the four sectors split into two subspaces: three with even fermion parity and one with odd fermion parity. Exactly which one corresponds to the odd fermion parity depends on the sign of $\mu$ and the parities of $L_x$ and $L_y$.  Interestingly, the $\mu>0$ and $\mu<0$ cases have distinct patterns of ground state fermion parities.  If $\mu<0$, the ground state fermion parity does not depend on the parities of $L_x$ and $L_y$, and the $\Phi_x=0,\Phi_y=0$ sector always has the odd fermion parity.

We now consider implementation of a Dehn twist, by shearing the lattice along the $y$ direction as depicted in \figref{fig:twist}(a). We weaken the bonds connecting $(0,y)$ and $(1,y)$, and at the same time strengthen the diagonal bonds connecting $(0,y)$ and $(1,y+1)$. This is done by modifying the Hamiltonian $H$ in \eqref{eq: H} to $H(\lambda)=H+\lambda H_1$ where
\begin{equation}
\begin{split}
H_1=&-\sum_{y}\big(c_{(0,y)}^\dag c_{(1,y+1)} -\Delta c_{(0,y)}^\dag c_{(1,y+1)}^\dag +\text{h.c.})\\
&+\sum_{y}\big( c_{(0,y)}^\dag c_{(1,y)} -\Delta c_{(0,y)}^\dag c_{(1,y)}^\dag +\text{h.c.} \big).
\end{split}
\label{}
\end{equation}
$\lambda$ going from $0$ to $1$ is one step of the Dehn twist. Repeating this procedure for $L_y$ steps accomplishes one full Dehn twist.   The flux configurations  are transformed from $(\Phi_x, \Phi_y)$ to $(\Phi_x+\Phi_y, \Phi_y)$. 

To have a well-defined Berry phase, we need to make sure that during the whole shearing process the system remains gapped (in the thermodynamic limit). To this end, notice that the modification Hamiltonian $H_1$ basically creates a one-dimensional channel along the $y$-direction, which can potentially localizes subgap states. More concretely, because during the entire twist the translation invariance along the $y$-direction is preserved, we can partially diagonalize the Hamiltonian by going to the Fourier space
$c_x(k_y)=L_y^{-1/2}\sum_y c_{(x,y)}e^{-\ii k_y y}$,
and obtain a collection of one-dimensional Hamiltonians along $x$ labeled by the momentum $k_y$. The fluxe $\Phi_y$ merely change the quantization condition of the momentum. It is easy to see that for the purpose of studying subgap states, we only need to consider $k_y=\pi$ and the 1D Hamiltonian reads
\begin{equation}
  \begin{split}
	H_{k_y=\pi}&=-\sum_{x\neq 0} (c_{x}^\dag c_{x+1}+\text{h.c.})+(2-\mu)\sum_x c_{x}^\dag c_{x}\\
	&+\Delta\sum_{x\neq 0}(c_{x}^\dag c_{x+1}^\dag+\text{h.c.})\\
	&+(2\lambda-1)(c_{0}^\dag c_{1}-\Delta c_{0}^\dag c_{1}^\dag+\text{h.c.})
  \end{split}
    \label{eqn:1d}
\end{equation}
Here $c_x\equiv c_x(\pi)$.
One immediately recognizes that this describes a one-dimensional spinless $p$-wave superconductor, belonging to symmetry class D in the topological superconductor periodic table.\cite{schnyder2008, kitaev2009}
If $0<\mu<4$, the one-dimensional model is in the nontrivial topological phase. As the boundary coupling $2\lambda-1$ changes sign, the ground state fermion parity of the one-dimensional Hamiltonian also changes and there must be a level crossing.\footnote{This is the essence of the $4\pi$ Josephson effect of class D topological superconductors.} As a result, gapless edge modes emerge along the twist line in the course of Dehn twist, which makes the evolution non-adiabatic in the thermodynamic limit.\footnote{Even in a finite system where one can suitably chose $L_y$ to avoid the $k_y=\pi$ point and maintain a finite-size gap during the whole twist, the subgap mode still has significant contribution to the Berry phase.}
On the other hand, when $\mu<0$ \eqref{eqn:1d} describes a trivial one-dimensional superconductor and there are no subgap modes for any value of $\lambda$, so the bulk gap remains finite. For $k_y=0$, the boundary coupling (between the $x=0$ and $x=1$ sites) remains a constant, so there can not be any subgap states. In conclusion, the lattice Dehn twist is an adiabatic evolution only when $\mu<0$. So in the following, we will set the chemical potential at $\mu=-2$, and study the Dehn twist on the $p_x+\ii p_y$ topological superconductor.

\emph{$S$ and $T$ Matrices.}--- 
We now present numerical results of the $T$ matrix for a $p_x+\ii p_y$ superconductor on a finite torus. We label the ground state in the flux sector $(\Phi_x, \Phi_y)=\pi(a, b)$ as $\ket{\Psi_{ab}}$ with $a,b=0,1$. In this basis, the $T_y$ matrix takes the following form:
\begin{equation}\label{eq: T}
T_y=
\begin{pmatrix}
e^{\ii\Theta_{00}} & 0 & 0 & 0\\
0 & 0 & e^{\ii\Theta_{01}} & 0 \\
0 & e^{\ii\Theta_{11}} & 0 & 0\\
0 & 0 & 0 & e^{\ii\Theta_{10}}
\end{pmatrix}.
\end{equation}
Schematically, $T_y\ket{\Psi_{ab}}=e^{\ii\Theta_{ab}}\ket{\Psi_{a+b\text{ mod }2,b}}$ transforms the fermion many-body state $\ket{\Psi_{ab}}$ to the new state $\ket{\Psi_{a+b\text{ mod }2,b}}$ evolved from the original state under a full Dehn twist along the $y$ direction. 

\begin{figure}[!t]
	\centering
	\includegraphics[width=\columnwidth]{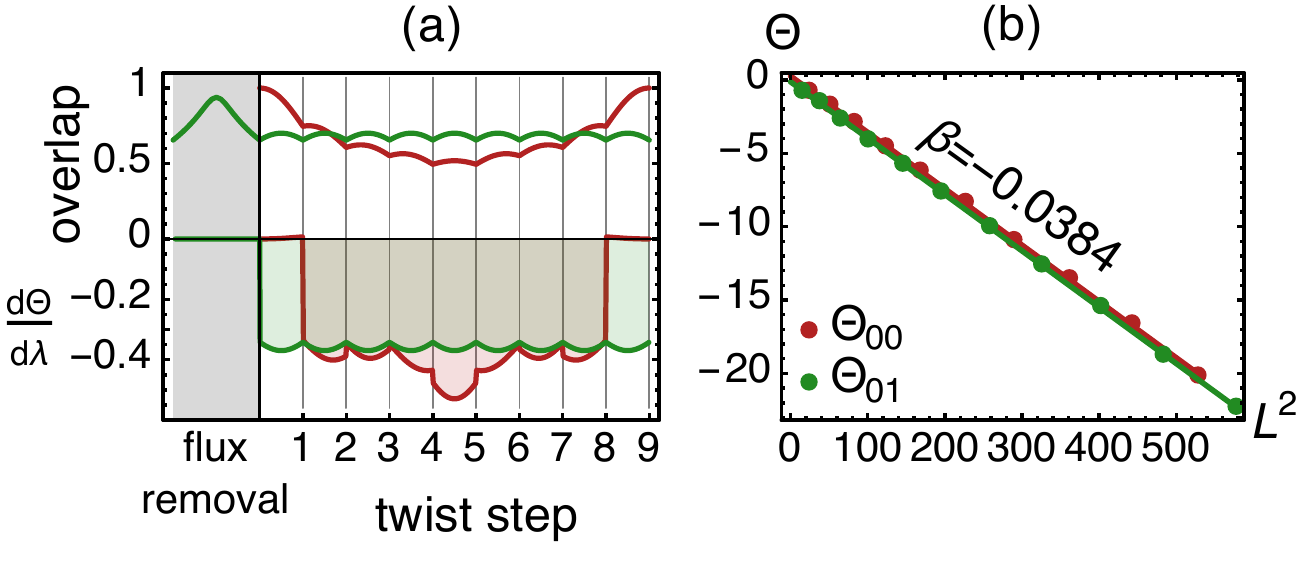}
	\caption{(a) The wave function overlap (upper panel) with the reference state and the Berry connection $\dd\Theta/\dd\lambda$ (lower panel) along the path of a full Dehn twist, with the initial flux configuration being $(0,0)$ (in red) or $(0,\pi)$ (in green). (b) The Dehn twist Berry phase $\Theta$ v.s. $L^2$, fitted by $\Theta=\alpha+\beta L^2$ for large-$L$ data to determine $\beta$ (which is the same for all flux configurations).}
	\label{fig:phase}
\end{figure}

\begin{figure}[t]
	\centering
	\includegraphics[width=0.92\columnwidth]{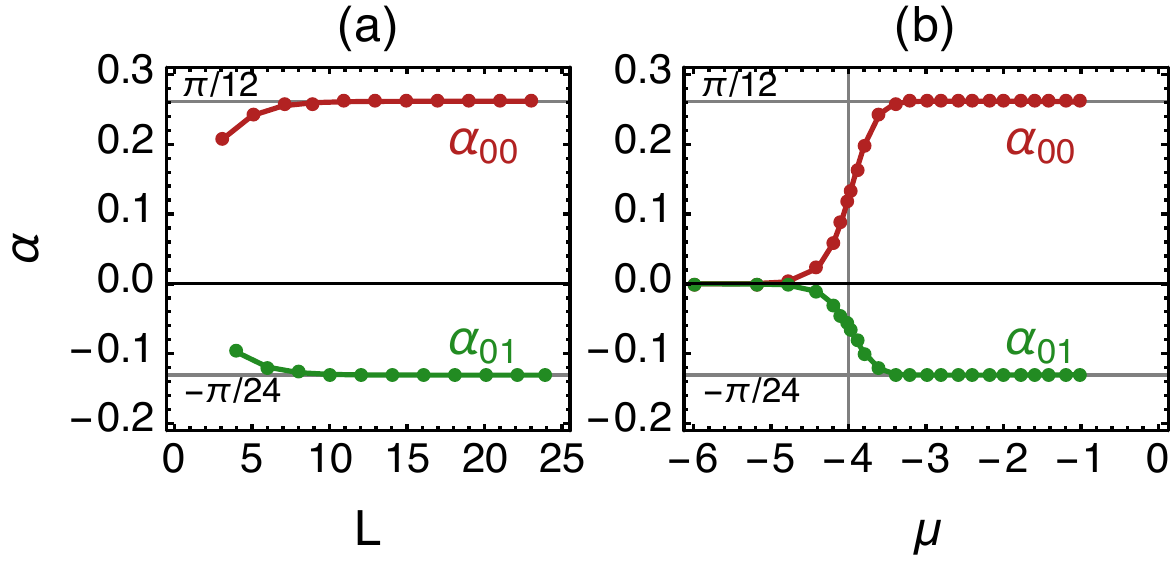}
	\caption{(a) The phase $\alpha$ v.s. system size $L$. $\alpha=\Theta-\beta L^2$ is obtained from the calculated Berry phase $\Theta$ subtracted by $\beta L^2$, where $\beta$ determined from the fitting of the large-$L$ data. (b) The jump of the universal phase $\alpha$ across the topological-trivial transition. $\mu>-4$ ($\mu<-4$) corresponds to the topological (trivial) phase. The calculation is performed on a $20\times20$ lattice. The sharp transition at $\mu=-4$ is smoothed out by the finite-size effect.}
	\label{fig:alpha}
\end{figure}

In \figref{fig:phase}(a) we show the accumulation of Berry phase during the evolution. In each step of the twist, $\lambda$ is smoothly tuned from 0 to 1. For an $L_x\times L_y$ sized lattice, we need $L_y$ steps of twists to complete a full Dehn twist. For example, the cases shown in \figref{fig:phase}(a) are of $L_y=9$. Let $\ket{\Psi(\lambda)}$ be the many-body ground state of the Hamiltonian $H(\lambda)$. The Berry connection $\dd\Theta/\dd\lambda$ can be calculated by comparing the phases of the nearby states $\ket{\Psi(\lambda)}$ and $\ket{\Psi(\lambda+\dd\lambda)}$ as $\dd\Theta=\arg\langle\Psi(\lambda)|\Psi(\lambda+\dd\lambda)\rangle$. To fix the gauge, we need to choose a reference state $\ket{\Psi_\text{ref}}$ and make sure that there is no relative phase between $\ket{\Psi(\lambda)}$ and $\ket{\Psi_\text{ref}}$ along the way of the twist, i.e. $\arg\langle\Psi(\lambda)|\Psi_\text{ref}\rangle=0$ (see Appendix \ref{sec:overlap} for the calculation of the overlap of two BCS wavefunctions and Appendix  \ref{sec:momentumtrans} for a more efficient algorithm in the momentum space). The upper panel of \figref{fig:phase}(a) shows the overlap of the ground states $\ket{\Psi(\lambda)}$ with the reference state $\ket{\Psi_\text{ref}}$, which never vanishes, verifying that the Dehn twist is indeed performed adiabatically. The lower panel of \figref{fig:phase}(a) shows the Berry connection through out the whole process, and the Berry phase can be obtained from the accumulation $\Theta=\oint \dd\Theta$. 

It is worth pointing out that the $T_y$ matrix is not completely diagonal: it leaves both the $(0,0)$ and $(\pi,0)$ flux configurations invariant, but takes the  $(0,\pi)$ configuration to $(\pi,\pi)$ and vice versa. In order to get a well-defined Berry phase in the latter case, we complete the cycle by first starting from the $(\pi,\pi)$ configuration and adiabatically turn off $\Phi_x$ on the boundary bounds without twisting the lattice until we reach the $(0,\pi)$ configuration, which is dubbed as a ``flux removal'' procedure; and then Dehn twisting the lattice back to the $(\pi,\pi)$ configuration. \figref{fig:phase}(a) shows the whole evolution for both the $(0,0)$ case (which does not involves the flux removal) and the $(0,\pi)$ case (which involves the flux removal). It is evident from the numerical result that the flux removal procedure does not give any contribution to the Berry phase. 

We carry out the calculation on a $L\times L$ sized lattice, and then perform a finite-size scaling of $\Theta$ to extract the universal phase $\alpha$. We notice that for all the configurations, the values of $\beta$ are almost identical (see \figref{fig:phase}(b)), so we can first determine $\beta$ from the fitting of the large-$L$ data, and then calculate $\alpha$ by subtracting $\beta L^2$ from the calculated $\Theta$, i.e. $\alpha=\Theta-\beta L^2$. \figref{fig:alpha}(a)  shows that $\alpha$ approaches to its universal value in the thermodynamic limit as $L\to\infty$. We notice that the value of $\beta$ is actually related to the Hall viscosity of the fluid once we take the continuum limit: 
$\beta=-\eta_H$ where $\eta_H\propto\bar{n}$ for $p_x+\ii p_y$ superfluid with $\bar{n}$ being the number density~\cite{AvronPRL1995, ReadPRB2009,ReadPRB2011, ZaletelPRL2013}. We confirm numerically the linear dependence of $\beta$ on $\bar{n}$ in the low-density regime (see Appendix \ref{sec:viscosity} for more details).

Up to numerical error, we find that $\alpha_{00}=\alpha_{10}=\pi/12$ and $\alpha_{01}=\alpha_{11}=-\pi/24$. Plugging into \eqref{eq: T}, the $T$ matrix is therefore
\begin{equation}\label{eq: Ty nu=1}
	T_y=e^{-\frac{\pi \ii}{24}}
\begin{pmatrix}
e^{\frac{\pi \ii}{8}} & 0 & 0 & 0\\
0 & 0 & 1 & 0 \\
0 & 1 & 0 & 0\\
0 & 0 & 0 & e^{\frac{\pi \ii}{8}}
\end{pmatrix}.
\end{equation}
Notice that we drop the overall non-universal phase $e^{\ii\beta L^2}$.
The overall phase $e^{-\frac{\pi \ii}{24}}=e^{-\frac{2\pi \ii}{24}\cdot\frac{1}{2}}$ is consistent with the chiral central charge $c_-=1/2$. We can also perform the Dehn twist by shearing along the $x$ direction, and calculate the corresponding $T$ matrix using the same method in the same set of basis. It is found that
\begin{equation}
T_x=e^{\frac{\pi \ii}{24}}
\begin{pmatrix}
e^{-\frac{\pi \ii}{8}} & 0 & 0 & 0\\
0 & e^{-\frac{\pi \ii}{8}} & 0 & 0 \\
0 & 0 & 0 & 1\\
0 & 0 & 1 & 0
\end{pmatrix}.
\end{equation}
Then the $S$ matrix can be obtained from $S=T_yT_x^{-1}T_y$, because the $S$ transform corresponds to a $\pi/2$ rotation, which can be composed from three shear transformations (Dehn twists):
\begin{equation}\label{eq: S nu=1}
S=T_yT_x^{-1}T_y=
\begin{pmatrix}
e^{\frac{\pi \ii}{4}} & 0 & 0 & 0\\
0 & 0 & 0 & 1 \\
0 & 0 & 1 & 0\\
0 & 1 & 0 & 0
\end{pmatrix}.
\end{equation}

If our model is viewed as fermions coupled to static $\mathbb{Z}_2$ gauge fields, the physical states must be gauge-invariant. So  it is necessary to project to even fermion parity subspace which corresponds to the lower $3\times 3$ block. In the $T_y$-diagonal basis, 
\begin{equation}
T_y= e^{-\frac{\pi \ii}{24}}
\begin{pmatrix}
1 & 0 & 0\\
0 & e^{\frac{\pi \ii}{8}} & 0\\
0 & 0 & -1
\end{pmatrix},
S=
\begin{pmatrix}
\frac{1}{2} & \frac{1}{\sqrt{2}} & \frac{1}{2}\\
\frac{1}{\sqrt{2}} & 0 & -\frac{1}{\sqrt{2}} \\
\frac{1}{2} & -\frac{1}{\sqrt{2}} & \frac{1}{2}
\end{pmatrix}.
\label{eq: Ising ST}
\end{equation}
As expected, they match the topological $T, S$ matrices of the Ising anyon model~\cite{Kitaev06a}. It is interesting to observe that although the $(0,0)$ sector is thrown away in the gauge projection (because its fermion parity is odd), the Berry phase of the Dehn twist is nevertheless still well-defined, which turns out to be the topological twist of the $\pi$-flux in the gauged model.

\emph{TQFT Interpretation.}---
The above numerical results can be interpreted from a TQFT perspective. As we have mentioned, our model can be considered as a $p_x+\ii p_y$ superconductor coupled to a static $\mathbb{Z}_2$ gauge field, and the TQFT description is an Ising anyon model~\cite{Kitaev06a}, which  has three topological charges, $1, \sigma$ and $\psi$. Here $\psi$ corresponds to the fermion in the lattice model, and $\sigma$ corresponds to the $\pi$ flux with a localized Majorana zero mode. They satisfy the well-known fusion rules:
\begin{equation}
	\sigma\times\sigma=1+\psi, \sigma\times\psi=\sigma.
	\label{}
\end{equation}
On a torus, the topologically-protected degenerate ground states can be labeled by the topological charge that going through one of the two holes, e.g. the longitudinal one (often refered to as an anyon flux), and a topological charge measurement can be performed along the meridian cycle to distinguish the different ground states. The $S$ and $T$ matrices can be understood as basis transformations on the torus.

\begin{figure}[t]
	\centering
	\includegraphics[width=0.65\columnwidth]{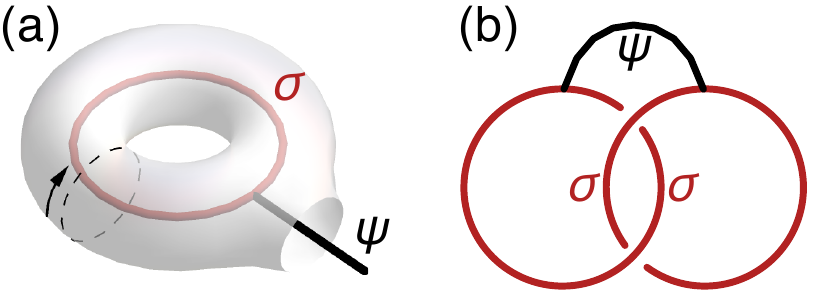}
	\caption{(a) The odd fermion parity sector can be view as a fermion charge line $\psi$ going out from the punctured torus.  (b) Diagrammatic definition of the $S$ matrix in the odd fermion parity sector.}
	\label{fig:puncture}
\end{figure}

For the $p_x+\ii p_y$ superconductor, the three ground states with even fermion parity are unitarily related to the states with definite topological charges $1, \sigma, \psi$. However, the odd fermion parity sector $(0,0)$ requires a different interpretation: formally this state does not exist on a closed surface since it carries an odd $\mathbb{Z}_2$ charge. Instead one has to use a punctured torus with a $\psi$ charge line coming out of the punctured hole, see \figref{fig:puncture}(a). Judging from the fusion rule, this means that there must be a $\sigma$ charge line going in the longitudinal direction (see \figref{fig:puncture}(a) for an illustration), which explains why the Berry phase of the Dehn twist on the $(0,0)$ sector is equal to the topological twist of $\sigma$. The $S$ matrix is now given by $S_{\sigma\sigma}^{\psi}$~\cite{Kitaev06a} as shown in \figref{fig:puncture}(b), and our numerical result indeed agrees with the diagrammatic result.
We can also understand this result intuitively as follows:  assuming there is $\mathrm{SO}(2)$ rotation symmetry, a single fermion acquires $-1$ phase under a full $2\pi$ rotation(i.e. the topological spin). Therefore a $\frac{\pi}{2}$ rotation would results in $e^{\ii\pi/4}$ phase, which is consistent with the numerical calculation.

\emph{Topological Phase Transition.}---
The Hamiltonian \eqref{eq: H} can also describe a trivial superconductor if the chemical potential $\mu$ lies outside the band, i.e. $|\mu|>4$, separated from the topological phase by a quantum phase transition. In the above calculation, we have always set $\mu=-2$ in the topological phase. Now we tune the chemical potential across the phase transition, and perform the same calculation of the Dehn twist Berry phase. Fig.\ref{fig:alpha}(b) shows that the universal phases $\alpha$ identically drop to zero across the topological-trivial transition as the chemical potential is tuned below $-4$. So if we start with the trivial superconductor, the $S$ and $T$ matrices will be
\begin{equation}
	T_y=
	\begin{pmatrix}
		1 & 0 & 0 & 0\\
		0 & 0 & 1 & 0\\
		0 & 1 & 0 & 0\\
		0 & 0 & 0 & 1
	\end{pmatrix},
	S=
	\begin{pmatrix}
		1 & 0 & 0 & 0\\
		0 & 0 & 0 & 1\\
		0 & 0 & 1 & 0\\
		0 & 1 & 0 & 0
	\end{pmatrix}.
	\label{eq: toric code ST}
\end{equation}
For the trivial superconductor, all the flux configurations lead to even fermion parity states, so the $S$ and $T$ matrices remain the same as \eqref{eq: toric code ST} after gauging the fermion parity, which can be transformed\cite{liu2013,He_arxiv} to the standard form of the $S$ and $T$ matrices of the $\mathbb{Z}_2$ toric code model.  The TQFT description of a $\mathbb{Z}_2$ gauged trivial superconductor is indeed a toric code theory, which has four topological charges $1$, $e$, $m$ and $\psi$, whose topological twists are given by the eigenvalues of $T_y$: $1, 1, 1, -1$. $\psi$ corresponds to the fermion in the lattice model, $m$ corresponds to the $\pi$ flux in the trivial superconductor, and $e$ is the bound state of $\psi$ and $m$.

\emph{TSC with Higher Chern Number.}---
A topological superconductor with Chern number $\nu>0$ can be considered as $\nu$ copies of $p_x+\ii p_y$ superconductors. Each copy will accumulate the same amount of Berry phase during the Dehn twist, so the $S$ and $T$ matrices simply follow from \eqref{eq: Ty nu=1} and \eqref{eq: S nu=1} with all the Berry phases multiplied by $\nu$ times,
\mycomment{
\begin{equation}
	T_y=e^{-\frac{\pi \ii\nu}{24}}
	\begin{pmatrix}
		e^{\frac{\pi \ii\nu}{8}} & 0 & 0 & 0\\
		0 & 0 & 1 & 0\\
		0 & 1 & 0 & 0\\
		0 & 0 & 0 & e^{\frac{\pi \ii\nu}{8}}
	\end{pmatrix},
	S=
	\begin{pmatrix}
		e^{\frac{\pi \ii\nu}{4}} & 0 & 0 & 0\\
		0 & 0 & 0 & 1\\
		0 & 0 & 1 & 0\\
		0 & 1 & 0 & 0
	\end{pmatrix}.
	\label{}
\end{equation}
}
For odd $\nu$, the even fermion parity sector corresponds to the lower $3\times3$ block; while for even $\nu$, all the four basis states are of even fermion parity. 

\emph{Conclusions and Discussions.}--- In this work, we calculated both the $S$ and $T$ modular matrices of two-dimensional topologically ordered phases that can be constructed by gauging the fermion parity from a free-fermion topological superconductors. The modular matrices can be obtained as the non-Abelian Berry phases among the ground states in different gauge flux sectors. We use Dehn twists along both the meridian and the longitudinal cycles of the torus as the generators of the modular group, which can be conveniently implemented by smoothly shearing the lattice in both directions. 
 It will be interesting to generalize this method to calculate the modular data of interacting topological phases on lattices such as fractional Chern insulators~\cite{ShengNC2011, RegnaultPRX2011, NeupertPRL2011}, as well as $G$-crossed modular data for $G$-symmetry-protected phases~\cite{BBCW}, which will be left for future investigations.

\emph{Acknowledgement.}--- We are grateful for Maissam Barkeshli, Roger Mong, Yizhi You and Zhenghan Wang for enlightening discussions.

\appendix

\section{Overlap of BCS wavefunctions}
\label{sec:overlap}
In this appendix we summarize the algorithm to calculate the overlap of BCS wavefunctions. Let us rewrite the complex fermions $c_\vr$ in terms of Majorana fermions $\chi_i$, s.t. $c_\vr=\chi_{\vr,1}+\ii\chi_{\vr,2}$. In general the BCS Hamiltonian can be written in the Majorana basis
\begin{equation}
H(\lambda)=\frac{\ii}{2}\sum_{i,j}\chi_i A_{ij}(\lambda)\chi_j,
\end{equation}
where $\lambda$ is an external parameter. Diagonalizing the Hermitian matrix $\ii A$, one finds its eigen states $u_n$ along with the eigen energies $E_n$: $\ii A u_n=E_n u_n $. We only consider the gapped fermion system, so the spectrum contains half of the states $u_n$ with negative energy $E_n<0$. The fermionic many-body ground state $\ket{\Psi}$ of $H$ is simply constructed by occupying all the negative energy single-particle states. Let us arrange the negative energy states $u_n$ (in the form of column vectors) into a matrix
\begin{equation}
U=\begin{pmatrix}\cdots & u_n & \cdots\end{pmatrix}_{E_n<0},
\end{equation}
such that each column of $U$ is a state vector $u_n$. Of course, this matrix $U(\lambda)$ also depends on the external parameter $\lambda$. Then the overlap of two BCS ground states $\ket{\Psi(\lambda_1)}$ and $\ket{\Psi(\lambda_2)}$ can be expressed in terms of the corresponding $U$ matrices as
\begin{equation}\label{eq: overlap}
\langle\Psi(\lambda_1)|\Psi(\lambda_2)\rangle=\pf\big[U^\dag(\lambda_1) U(\lambda_2)\big].
\end{equation}
With this, one can calculate the overlap between any pair of BSC ground states.

However the eigen states $u_n$ obtained from diagonalization of $A$ is subject to an overall phase degree of freedom, and hence the $U$ matrix is also ambiguous. To remove this phase ambiguity and fixed the gauge, we need to choose a reference state $\ket{\Psi_\text{ref}}$, so that we can define the Berry phase accumulated from $\lambda$ to $\lambda+\dd\lambda$ to be
\begin{equation}
\begin{split}
\dd\Theta =& \arg\langle\Psi(\lambda)|\Psi(\lambda+\dd\lambda)\rangle\\
&-\arg\langle\Psi(\lambda)|\Psi_\text{ref}\rangle+\arg\langle\Psi(\lambda+\dd\lambda)|\Psi_\text{ref}\rangle.
\end{split}\label{eq: Berry}
\end{equation}
Alternatively, one may first rectify the overall phase of $\ket{\Psi(\lambda)}$ with respect to the reference state, such that $\arg\langle\Psi(\lambda)|\Psi_\text{ref}\rangle=0$ for all $\lambda$, then \eqref{eq: Berry} is reduced to $\dd\Theta = \arg\langle\Psi(\lambda)|\Psi(\lambda+\dd\lambda)\rangle$. On every given small segment $\dd\lambda$, $\dd\Theta$ may vary with the choice of the reference state $\ket{\Psi_\text{ref}}$. But after integrating over a closed path, the Berry phase $\Theta=\oint \dd \Theta$ is well-defined and independent of $\ket{\Psi_\text{ref}}$. In the calculation of the Berry phase, it is import to check that the overlap to the reference state $|\langle\Psi(\lambda)|\Psi_\text{ref}\rangle|$ never becomes vanishingly small, otherwise the result will suffer from severe numerical error.  

In our calculation, for $\alpha_{00}$ and $\alpha_{10}$, the reference state is chosen to be the initial state $\ket{\Psi_\text{ref}}=\ket{\Psi(\lambda=0)}$. While for $\alpha_{01}$ and $\alpha_{11}$, the reference state is chosen to be the ground state of $H$ with the boundary condition being open in the $x$-direction and anti-periodic in the $y$-direction.

\section{Dehn Twist in Momentum Space}
\label{sec:momentumtrans}
The lattice Hamiltonian in \eqref{eq: H} can be easily diagonalized in the momentum space as \eqref{eq: H in k}. On the twisted lattice, the translational symmetry that is perpendicular to the twist should be understood as the translation followed by a shearing along the Dehn twist direction, therefore the quasi-momenta (eigenvalues of the translation symmetry) are quantized to (on an $L\times L$ square lattice)
\begin{equation}
\vk=\frac{1}{L}\begin{pmatrix}1 & 0 \\ -\tau_y & 1 \end{pmatrix}
\begin{pmatrix} 2\pi n_x+\Phi_x \\ 2\pi n_y+\Phi_y\end{pmatrix},\quad n_x,n_y\in\mathbb{Z},
\end{equation}
where $\tau_y$ evolves from 0 to 1 corresponding to a full Dehn twist along the $y$-direction, and $(\Phi_x,\Phi_y)$ denotes the initial flux configuration (at the beginning $\tau_y=0$ of the Dehn twist).

For each $\vk$ in the Brillouin zone, we can simply diagonalize the $2\times2$ single-particle Hamiltonian $H_\vk$ given in \eqref{eq: H in k} to obtain the Bloch state vector $u_\vk$, i.e. $H_\vk u_\vk = E_\vk u_\vk$ (with $E_\vk<0$). Then each occupied single-particle state is labeled by both its quasi-momentum $\vk$ and the corresponding Bloch state vector $u_\vk$, denoted as $\ket{\vk,u_\vk}$. The overlap between two single-particle states is given by
\begin{equation}
\langle \vk,u_\vk|\vk',u_{\vk'}\rangle=\frac{\sin\big[\frac{1}{2}(\vk-\vk') L\big]}{L\sin\big[\frac{1}{2}(\vk-\vk')\big]} u_{\vk}^\dagger u_{\vk'}.
\end{equation}
With this, the overlap of two fermionic many-body states that are parametrized by $\tau_{y,1}$ and $\tau_{y,2}$ follows from \eqref{eq: overlap} as
\begin{equation}
\langle\Psi(\tau_{y,1})|\Psi(\tau_{y,2})\rangle =\pf\big[{}_{\tau_{y,1}}\langle \vk,u_\vk|\vk',u_{\vk'}\rangle_{\tau_{y,2}}\big],
\end{equation}
where $\ket{\vk,u_{\vk}}_{\tau_y}$ denotes the single-particle state that is obtained from the Hamiltonian at the parameter $\tau_y$, and the Pfaffian matrix index runs over $\vk$ (row index) and $\vk'$ (column index) through out the Brillouin zone. Then the calculation of the Berry phase is similar to the real space algorithm given in \eqref{eq: Berry}. In this momentum space representation, the flux insertion/removal procedure can be simply implemented by smoothly tuning $(\Phi_x,\Phi_y)$. The computation efficiency can be improved significantly using the momentum space formalism, because the single-particle Hamiltonian is very easy to diagonalize in the momentum space.

\section{Hall viscosity from the Lattice Model}
\label{sec:viscosity}
We change the number density of the $p_x+i p_y$ superfluid by tuning the chemical potential $\mu$ in the model Hamiltonian \eqref{eq: H}. For each density $\bar{n}$, we calculated the Dehn twist Berry phase $\Theta=\alpha+\beta L^2+O(L^{-2})$ for various $L$, and extract the coefficient $\beta$. $\beta$ is related to the Hall viscosity $\eta_H$ of the fluid in the continuum limit: $\beta=-{\eta_H}$ which is proportional to the number density $\bar{n}$. It is verified that $\beta$ indeed scales linearly with $\bar{n}$ in the dilute limit $\bar{n}\to0$, as expected.

\begin{figure}[htbp]
\begin{center}
\includegraphics[width=136pt]{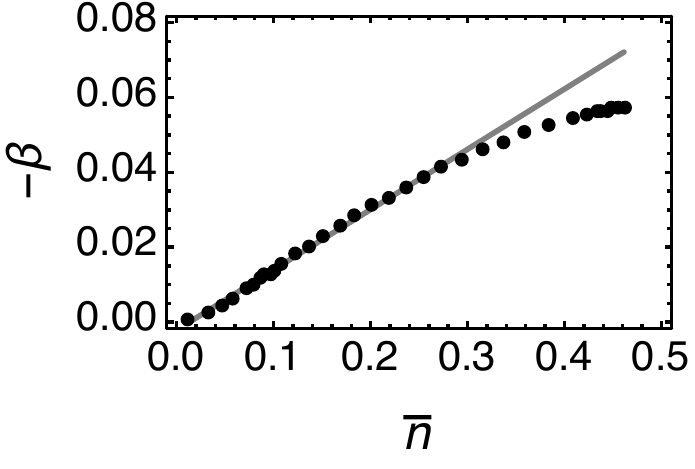}
\caption{Hall viscosity scales with the number density $\bar{n}$ linearly in the dilute limit $\bar{n}\to0$.}
\label{fig:viscosity}
\end{center}
\end{figure}

\bibliography{corr}

\end{document}